\documentclass[reprint,
showpacs,
 amsmath,amssymb,
prb,
]{revtex4-1}

\usepackage{graphicx}
\usepackage{dcolumn}
\usepackage{longtable}
\usepackage{lipsum}
\usepackage{color}

\begin{document}


\title{ Odd-frequency Cooper pairs in two-band superconductors and their magnetic response}
\author{Yasuhiro Asano$^{1,2,3}$}
\author{Akihiro Sasaki$^{1}$}
\affiliation{$^{1}$Department of Applied Physics,
Hokkaido University, Sapporo 060-8628, Japan\\
$^{2}$Center of Topological Science and Technology,
Hokkaido University, Sapporo 060-8628, Japan\\
$^{3}$Moscow Institute of Physics and Technology, 141700 Dolgoprudny, Russia}%

\date{\today}

\begin{abstract}
We discuss the appearance of odd-frequency Cooper pairs  
in two-band superconductors by solving the Gor'kov equation 
analytically. We introduce the equal-time $s$-wave pair potentials
as realized in MgB$_2$ and iron pnictides. 
Although the order parameter symmetry is conventional,
the band degree of freedom enriches the symmetry variety of pairing correlations.  
The hybridization and the asymmetry between the two conduction bands 
induce odd-frequency pairs as a subdominant pairing correlation 
in the uniform ground state. To study the magnetic response of odd-frequency Cooper 
pairs, we analyze the Meissner kernel represented by the Gor'kov Green function. 
In contrast to the even-frequency pairs linked to the pair potential,
the induced odd-frequency Cooper pairs indicate a paramagnetic property.
We also discuss the relation between the amplitude of the odd-frequency pairing correlation   
and the stability of superconducting states in terms of the self-consistent equation 
for the pair potential.
\end{abstract}


\maketitle

\section{Introduction}
The electronic structure at the Fermi level governs such characteristics of 
superconductivity as effective dimensionality, 
anisotropy in electromagnetic properties and the symmetry of the superconducting order 
parameter. Many of the superconductors discovered so far have multibands at 
the Fermi level. These indicate unique characteristics such as 
a high-critical transition temperature in iron pnictides~\cite{pnictide},
unconventional superconductivity in heavy fermionic compounds~\cite{heavy1,heavy2,heavy3},
unusual vortex states~\cite{moshchalkov} in MgB$_2$~\cite{mgb2,mgb22}, and 
topologically nontrivial superconducting states in Cu-doped Bi$_2$Se$_3$~\cite{hor,fu}.
A microscopic understanding of the pairing mechanism would make it possible to explain these features.

At the superconducting transition temperature $T_c$, a superconductor chooses three discrete
symmetry options for Cooper pairing: frequency symmetry,
spin configuration, and momentum parity. In each case, there are two
possibilities: pairing can be either symmetric or antisymmetric with
respect to the interchange of the corresponding arguments, namely the times,
spins, or coordinates of the two electrons forming a Cooper pair.
Black-Schaffer and Balatsky~\cite{black-schaffer} have shown that
the multiband superconducting order parameter has an extra alternative symmetry 
option that 
originates from the band degree of freedom, namely even-band-parity and odd-band-parity. 
The Fermi-Dirac statistics of electrons requires
the constraint that the pairing functions must be antisymmetric
under the permutation of the two electrons.
As a consequence, Cooper pairs can be classified into eight 
symmetry classes as shown in Table~\ref{table1}. 
The classification of Cooper pairs in single-band superconductors corresponds to 
the top four classes in Table~\ref{table1}. 
Conventional superconductors and $d$-wave high-$T_c$ superconductors 
belong to the ESEE class. Spin-triplet $p$-wave superconductors such as Sr$_2$RuO$_4$ and UPt$_3$ 
belong to the ETOE class. It is known that the spatial inhomogeneity of the pair potential 
in the ETOE (ESEE) superconductors generates 
Cooper pairs that belong to the OTEE (OSOE) class~\cite{yt07,review,diaodd}.
Actually odd-frequency Cooper pairs~\cite{berezinskii} appear 
at a surface of unconventional superconductors as a subdominant pairing correlation~\cite{yt07,review,diaodd}.
The possibilities of odd-frequency superconductivity  
have also been discussed theoretically~\cite{balatsky92,vojta,kirkpatrik,coleman,solenov,kusunose,hoshino}.
Recent theoretical papers studied the mechanisms of odd-frequency superconductivity in multiband 
superconductors~\cite{komendova,aperis}.
The bottom four symmetry classes have been pointed out by the authors of Ref.~\onlinecite{black-schaffer}.
Indeed, they show that the hybridization between the two conduction bands generates Cooper pairs 
belonging to the OSEO class. 
The appearance of odd-frequency pairs in a uniform ground state may be a surprising conclusion 
when we pay attention to a unique property of them.

Although diamagnetism is the most fundamental property of all superconductors,
a number of theoretical studies have suggested 
that odd-frequency Cooper pairs have a \textsl{paramagnetic} property~\cite{yt05r,ya11,diaodd,higashitani1,mironov}.
A $\mu$SR experiment~\cite{bernardo1} has caught a clear sign of paramagnetic Cooper pairs very recently.
In addition, a recent experimental finding of a zero-bias anomaly in a ferromagnet/superconductor proximity 
structure~\cite{bernardo2}, which is also closely related to the paramagnetic property, has suggested 
the existence of odd-frequency pairs~\cite{bergeret,yt04,tshape}.
The paramagnetic Cooper pairs attract magnetic fields~\cite{suzuki1,suzuki2}, which implies 
the thermodynamic instability of odd-frequency pairing states. (See also Appendix A for details.)
Actually, in single-band superconductors, 
odd-frequency pairs always appear as a spatially localizing subdominant pairing correlation. 
Thus the instability of an odd-frequency pair does not affect the transition temperature 
if the superconductor is sufficiently large.
However, we have never encountered the type of magnetic response exhibited by induced odd-frequency pairs
in multiband superconductors. If they are paramagnetic, the superconducting condensate 
might be unstable due to the presence of odd-frequency pairs.
We address these issues in the present paper.

In this paper, we will show the existence of odd-frequency pairs in a two-band superconductor 
with the spin-orbit interactions or the band asymmetry. 
Within the theoretical model that we consider in this paper, odd-frequency Cooper pairs always 
exhibit a 
paramagnetic property irrespective of their generation mechanism. 
Since odd-frequency pairs appear as a subdominant pairing correlation in 
a uniform ground state, 
their paramagnetic instability suppresses the transition temperature.
We also find that the band hybridization generates even-frequency Cooper pairs 
whose symmetry is different from that of the order parameter.
We conclude that the rich symmetry variety of the induced pairing correlations in the bulk 
state is a key feature of multiband superconductors.

This paper is organized as follows. In Sec.~II, we present a theoretical model 
describing two-band superconductors.
The anomalous Green function of the Gor'kov equation 
are calculated and the magnetic property of odd-frequency pairs is discussed 
for the intra- and interband pairing order parameter in Sec.~III and 
 Sec.~IV, respectively. 
The conclusion is provided in Sec.~V.

\begin{table}[t]
\begin{center}
\begin{ruledtabular}
\begin{tabular}{lcccc}
\null &Frequency  & Spin & Momentum parity & Band parity\\
\hline
ESEE &  Even & Singlet & Even & Even \\
ETOE &  Even & Triplet & Odd & Even \\
OSOE &   Odd & Singlet & Odd & Even \\
OTEE &   Odd & Triplet & Even & Even \\
\hline
ESOO &  Even & Singlet & Odd & Odd \\
ETEO &  Even & Triplet & Even & Odd \\
OSEO &   Odd & Singlet & Even & Odd \\
OTOO &   Odd & Triplet & Odd & Odd \\
\end{tabular}
\end{ruledtabular}
\caption{
Eight symmetry classes of Cooper pairs in the presence of the band degree of freedom.
}
\end{center}
\label{table1}
\end{table}
%
\section{Two-band model}
The Hamiltonian of two-band electronic states is,
\begin{align}
&H_{\mathrm{N}}=\sum_{\alpha=\uparrow, \downarrow} \sum_{\lambda =1, 2}
\int d\boldsymbol{r} \psi^\dagger_{\lambda,\alpha}(\boldsymbol{r}) 
\xi_\lambda(\boldsymbol{r}) \psi_{\lambda,\alpha}(\boldsymbol{r}) \nonumber\\
+&\sum_{\alpha}
\int d\boldsymbol{r} \left\{
\psi^\dagger_{1,\alpha}(\boldsymbol{r}) V_0(\boldsymbol{r})
\psi_{2,\alpha}(\boldsymbol{r}) 
+ \psi^\dagger_{2,\alpha}(\boldsymbol{r}) V^\ast_0(\boldsymbol{r})
\psi_{1,\alpha}(\boldsymbol{r}) \right\} \nonumber\\
+&\sum_{\alpha,\beta}
\int d\boldsymbol{r} \left\{
\psi^\dagger_{1,\alpha}(\boldsymbol{r}) (i\boldsymbol{L}\times \nabla )\cdot \boldsymbol{\sigma}_{\alpha,\beta}
\psi_{2,\beta}(\boldsymbol{r}) \right.\nonumber\\
+&\left. \psi^\dagger_{2,\alpha}(\boldsymbol{r}) (i\boldsymbol{L}\times \nabla )\cdot \boldsymbol{\sigma}_{\alpha,\beta}
\psi_{1,\beta}(\boldsymbol{r}) \right\},
\end{align}
where $\xi_\lambda(\boldsymbol{r}) =- \frac{ \nabla^2}{2m_\lambda}-\mu_\lambda$,
$\hat{\sigma}_j$ for $j=1-3$ are the Pauli matrices in spin space, $\psi^\dagger_{\lambda,\alpha}(\boldsymbol{r})$
($\psi_{\lambda,\alpha}(\boldsymbol{r})$) is the creation (annihilation) operator of an electron 
in band $\lambda$ (1 or 2) with spin $\alpha$ ($\uparrow$ or $\downarrow$) and at a location $\boldsymbol{r}$.
The hybridization potential expressed by $V_0$ is independent of electron spin. 
We also consider the hybridization caused by the spin-orbit interaction represented by 
$\boldsymbol{L}$.

Throughout this paper, we assume the equal-time $s$-wave pair potential, 
(i.e., even-frequency even-momentum-parity class) described by
\begin{align}
\Delta_{\lambda \nu; \alpha \beta} =& g_{\lambda, \nu} 
\sum_{\boldsymbol{k}}
\left\langle \psi_{\lambda, \alpha}(\boldsymbol{k}) 
\psi_{\nu, \beta}(-\boldsymbol{k}) \right\rangle=- \Delta_{\nu\lambda; \beta \alpha},
\end{align} 
where $g_{\lambda, \nu}=g_{\nu, \lambda}$ represents the attractive 
interaction and $\langle \cdots \rangle$ represents the quantum and thermal average.
The last equation is derived from the Fermi-Dirac statistics of electrons.
In this paper, we confine ourselves to the pair potentials consisting of two electrons with opposite spin.
We define the spin-singlet intraband pair potential by
\begin{align}
\Delta_\lambda\equiv \Delta_{\lambda \lambda; \uparrow \downarrow}=- \Delta_{\lambda \lambda; \downarrow \uparrow},
\end{align}
which belongs to the ESEE class in Table~\ref{table1}. Such intraband pair potentials have been considered 
in previous literature~\cite{suhl,sung,golubov}.
There are two types of pair potentials for the interband order parameters. 
The pair potential is defined by
\begin{align}
\Delta_{12}\equiv &\Delta_{12; \uparrow \downarrow}=-\Delta_{21;  \downarrow \uparrow},\\
=& - s_{\mathrm{spin}} \Delta_{21; \uparrow \downarrow}= s_{\mathrm{spin}} \Delta_{12;  \downarrow \uparrow},
\end{align}
where $s_{\mathrm{spin}}$ is 1 for spin-triplet symmetry belonging to the ETEO class 
and -1 for spin-singlet symmetry belonging to the ESEE class.
\begin{widetext}

The BCS Hamiltonian in momentum space is represented by,
\begin{align}
H_{\textrm{BCS}}(\boldsymbol{k}) 
=\frac{1}{2}\left[
\begin{array}{cccccccc}
\xi_1 & 0 & V_0+V_3 & V_1-iV_2 & 0 & \Delta_1 & 0 & \Delta_{12}\\
0 &\xi_1 & V_1+iV_2 & V_0 -V_3 & -\Delta_1 & 0 &s_{\mathrm{spin}}\Delta_{12}& 0\\
V_0^\ast+V_3 & V_1-iV_2 & \xi_2 & 0& 0 & -s_{\mathrm{spin}}\Delta_{12} & 0 & \Delta_2 \\
V_1+iV_2 & V_0^\ast -V_3 & 0 & \xi_2 & -\Delta_{12} & 0
& -\Delta_2 & 0\\
0&-\Delta_1^\ast & 0 & -\Delta_{12}^\ast &-\xi_1 & 0 &
-V_0^\ast+V_3 & V_1+iV_2 \\
\Delta_1^\ast & 0& -s_{\mathrm{spin}}\Delta_{12}^\ast &0 & 0 &-\xi_1& V_1-iV_2 
& -V_0^\ast-V_3\\
0 & s_{\mathrm{spin}}\Delta_{12}^\ast& 0& -\Delta_2^\ast & -V_0+V_3 & V_1+iV_2 &-\xi_2 & 0 \\
\Delta_{12}^\ast &0 & \Delta_2^\ast & 0 & V_1-iV_2 &
-V_0-V_3&0&-\xi_2
\end{array}\right],\label{bcs88}
\end{align}
\end{widetext}
where
$\xi_{\lambda,\boldsymbol{k}} =\frac{ \boldsymbol{k}^2 }{2m_\lambda}-\mu_\lambda$, 
$V_1(\boldsymbol{k} )= - L_2 k_3 + L_3 k_2$, $
V_2(\boldsymbol{k})= -L_3  k_1 +L_1 k_3$, $ and
V_3(\boldsymbol{k})= -L_1 k_2 +L_2 k_1$.
We note that 
$V_j(\boldsymbol{k})$ for $j=1-3$ are odd-momentum-parity functions 
satisfying $V_j(\boldsymbol{k})=-V_j(-\boldsymbol{k})$.
In this paper, we assume that the hybridization potentials are much smaller 
than the Fermi energy in the two conduction bands. The basic property 
of the superconducting states are determined by the band structures and the effective attractive interaction
in the absence of hybridization. We will discuss the effects of weak hybridizations on superconducting
states.
In what follow, we calculate the Green functions
by solving the Gor'kov equation for Eq.~(\ref{bcs88}). 
To obtain the analytical expression of the Green function, we reduce 
the $8 \times 8$ particle-hole space in Eq.~(\ref{bcs88}) to 
two $4\times 4$ particle-hole spaces, $\mathcal{N}_1$ and $\mathcal{N}_2$, 
by selecting one hybridization potential among $V_j$ for $j=0-3$.

For practical calculations, we often consider a simple two-band model
described by
\begin{align}
m_1=& m_2=m, \quad
\mu_{1(2)}= \mu_F +(-) \delta \mu, \label{smodel}
\end{align}
with $\delta\mu \ll \mu_F$.

\section{intraband pairing order}
We first consider a two-band superconductor with equal-time 
spin-singlet $s$-wave intraband pair potentials (ESEE).
In Eq.~(\ref{bcs88}), we set $\Delta_{12}=0$.

\subsection{Spin-independent hybridization}
When spin-flip hybridization is absent (i.e.,$V_1=V_2=0$), Eq.~(\ref{bcs88}) can be block diagonalized.
The reduced Hamiltonian is represented by
\begin{align}
\check{H}_1(\boldsymbol{k})=&\left[
\begin{array}{cccc}
\xi_1 & W & s \Delta_1 & 0\\
W^\ast & \xi_2 & 0 & s\Delta_2 \\
s\Delta_1^\ast & 0& -\xi_1 & -W^\ast\\
0& s\Delta_2^\ast & -W& -\xi_2 
\end{array}
\right], \label{h1}\\
W=&V_0 + sV_3(\boldsymbol{k}).
\end{align}
In $\mathcal{N}_1$, the spin of an electron is $\uparrow$ and that of a hole is $\downarrow$. 
While in the other particle-hole space $\mathcal{N}_2$, the spin of an electron (a hole) 
is $\downarrow$ ($\uparrow$). 
We set the sign factor $s$ as $s=1$ in $\mathcal{N}_1$ and -1 in $\mathcal{N}_2$.

\begin{table*}[t]
\begin{center}
\begin{ruledtabular}
\begin{tabular}{lccccc}
\null & Frequency  & Spin & Momentum parity & Band parity & Magnetic response \\
\hline
Pair potential &  Even & Singlet & Even & Even (intra) & Diamagnetic \\
\hline
Induced by $V_0$ &  Even & Singlet & Even & Even (inter) & Diamagnetic \\
\null &   Odd & Singlet & Even & Odd  & Paramagnetic \\
\hline
Induced by $V_1$, $V_2$, $V_3$ &  Even & Triplet & Odd & Even (inter) & Diamagnetic \\
\null &  Odd & Triplet & Odd & Odd  & Paramagnetic 
\end{tabular}
\end{ruledtabular}
\caption{
The symmetry classification of Cooper pairs with the equal-time spin-singlet even-momentum-parity 
intraband (even-band-parity) pair potential.
}
\label{table2}
\end{center}
\end{table*}
The Green function is defined by
\begin{align}
&\left[ i\omega_n \check{1}- \check{H}(\boldsymbol{k}) \right] \check{G}(\boldsymbol{k},i\omega_n)
=\check{1},\label{gorkov}\\
&\check{G}(\boldsymbol{k},i\omega_n)
=\left[ \begin{array}{cc}
\hat{\mathcal{G}} & \hat{\mathcal{F}} \\
\underline{\hat{\mathcal{F}}} & \underline{\hat{\mathcal{G}}} \end{array}\right]_{(\boldsymbol{k},i\omega_n)}.
\end{align}
At $V_3=0$,
the anomalous Green function is calculated as
\begin{align}
\hat{\mathcal{F}}_1&(\boldsymbol{k},i\omega_n)
=\frac{s}{2Z_1} 
\left[
\left\{ (-X_1 +v_2^2) \Delta_+ +  ( K  + i v_1v_2) \Delta_-  \right\} \hat{\rho}_0 \right. \nonumber\\
&+ \left\{ v_1( \xi_+ \Delta_+ -\xi_-\Delta_- ) -iv_2 ( \xi_+ \Delta_- -\xi_-\Delta_+ ) \right\}\hat{\rho}_1 \nonumber\\
&+ \omega_n ( v_1  \Delta_- -i v_2 \Delta_+ ) \hat{\rho}_2 \label{f_tra}\\
&+ \left.\left\{ (-X_1 +v_1^2)  \Delta_- + ( K -iv_1v_2) \Delta_+  \right\} \hat{\rho}_3   \right], 
\nonumber
\end{align}
with
\begin{align}
Z_1=& X^2_1 - Y_1,\\
X_1 =& \frac{1}{2}[\omega_n^2 + \xi_+^2 + \xi_-^2 + |\Delta_+|^2 + |\Delta_-|^2 + |V_0|^2],\\
Y_1=& K^2 +|V_0|^2\xi_+^2 + v_1^2|\Delta_-|^2+ v_2^2|\Delta_+|^2- i2 v_1v_2 D_-, \label{y1}
\end{align}
where we have defined 
\begin{align}
\xi_+=&\frac{\xi_{1,\boldsymbol{k}} + \xi_{2,\boldsymbol{k}} }{2}, \quad
\xi_-=\frac{\xi_{1,\boldsymbol{k}} - \xi_{2,\boldsymbol{k}} }{2}, \\
\Delta_+=&\frac{\Delta_{1} + \Delta_{2} }{2}, \quad
\Delta_-=\frac{\Delta_{1} - \Delta_{2} }{2},\\
K=& \xi_+ \xi_- + D_+, \quad D_\pm= \frac{\Delta_+ \Delta_-^\ast \pm \Delta_+^\ast \Delta_-}{2},\label{k_def}\\
V_0=&v_1+iv_2.
\end{align}
Here $\hat{\rho}_j$ for $j=1-3$ are Pauli matrices in the two-band space.
The normal Green function is shown in Eq.~(\ref{g1}) in Appendix B.
The relation
\begin{align}
\underline{\hat{\mathcal{F}}}_1(\boldsymbol{k},i\omega_n)
=& \hat{\mathcal{F}}^\ast_1(\boldsymbol{k},i\omega_n).\label{ff1}
\end{align}
holds due to the symmetry of the Gor'kov equation for reduced Hamiltonian in Eq.~(\ref{h1}). 
The components of the anomalous Green function in Eq.~(\ref{f_tra}) are represented by 
\begin{align}
\hat{\mathcal{F}}_1^{\mathcal{N}_1}
=\left[\begin{array}{cc} \mathcal{F}_{11,\uparrow\downarrow} & \mathcal{F}_{12,\uparrow\downarrow} \\
\mathcal{F}_{21,\uparrow\downarrow} & \mathcal{F}_{22,\uparrow\downarrow}
\end{array}\right], \;
\hat{\mathcal{F}}_1^{\mathcal{N}_2}
=\left[\begin{array}{cc} \mathcal{F}_{11,\downarrow\uparrow} & \mathcal{F}_{12,\downarrow\uparrow} \\
\mathcal{F}_{21,\downarrow\uparrow} & \mathcal{F}_{22,\downarrow\uparrow}
\end{array}\right], \nonumber
\end{align}
in $\mathcal{N}_1$ and $\mathcal{N}_2$, respectively.
The Fermi-Dirac statistics of electrons requires 
\begin{align}
\mathcal{F}_{\lambda\lambda';\sigma\sigma'}(\boldsymbol{k},i\omega_n)=-
\mathcal{F}_{\lambda'\lambda;\sigma'\sigma}(-\boldsymbol{k},-i\omega_n). \label{antisymmetric}
\end{align}
The intraband components are represented as
\begin{align}
&\mathcal{F}_{11,\uparrow\downarrow}(\boldsymbol{k},i\omega_n)=
-\mathcal{F}_{11,\downarrow\uparrow}(\boldsymbol{k},i\omega_n) \nonumber\\
=& 
\frac{1}{2Z_1}\left[ -(X_1-K)(\Delta_+ +\Delta_-) + v_2^2\Delta_+ +v_1^2\Delta_- \right.\nonumber\\
&\left.- iv_1v_2(\Delta_+-\Delta_-) \right],\\
&\mathcal{F}_{22,\uparrow\downarrow}(\boldsymbol{k},i\omega_n)=-\mathcal{F}_{22,\downarrow\uparrow}(\boldsymbol{k},i\omega_n)
\nonumber\\
=& 
\frac{1}{2Z_1}\left[ -(X_1 +K)(\Delta_+ -\Delta_-) + v_2^2\Delta_+ -v_1^2\Delta_- \right. \nonumber\\
&+ \left.iv_1v_2(\Delta_++\Delta_-) \right].
\end{align}
These components are linked to the pair potential belonging to the ESEE class. 
The hybridization generates two types of interband pairing correlations.
The interband components result in
\begin{align}
&\mathcal{F}_{12,\uparrow\downarrow}(\boldsymbol{k},i\omega_n)=-\mathcal{F}_{12,\downarrow\uparrow}(\boldsymbol{k},i\omega_n)
=P_{1e} - P_{1o},\\
&\mathcal{F}_{21,\uparrow\downarrow}(\boldsymbol{k},i\omega_n)=-\mathcal{F}_{21,\downarrow\uparrow}(\boldsymbol{k},i\omega_n)
=P_{1e} + P_{1o},\\
&P_{1e} = \frac{ v_1(\xi_+\Delta_+-\xi_-\Delta_-) - iv_2(\xi_+\Delta_- -\xi_-\Delta_+ )}{2Z_1},\\
&P_{1o} = \frac{1}{2Z_1}\left[ i\omega_n(v_1\Delta_- - iv_2 \Delta_+) \right].
\end{align}
We find that the component
\begin{align}
F_{12,\uparrow\downarrow}-F_{12,\downarrow\uparrow}
+F_{21,\uparrow\downarrow}-F_{21,\downarrow\uparrow} = 4P_{1e},
\end{align}
belongs to the ESEE class and
\begin{align}
F_{12,\uparrow\downarrow}-F_{12,\downarrow\uparrow}
-F_{21,\uparrow\downarrow}+F_{21,\downarrow\uparrow} = -4P_{1o}, \label{f1_odd}
\end{align}
has the odd-frequency spin-singlet odd-band parity symmetry (OSEO).
The symmetry classification results are summarized in Table~\ref{table2}.
In Ref.~\onlinecite{black-schaffer}, the authors assume that $V_0$ is a real potential. 
In such a case, the odd-frequency pairing correlation appears only for $\Delta_-\neq 0$ 
because $v_2=0$. In the normal state, it is possible to eliminate 
the phase of the hybridization potential. Namely the phase factor $e^{i\phi}$ in
\begin{align}
\sum_{\alpha}& \int d\boldsymbol{r} \left[\psi_{1\alpha}^\dagger(\boldsymbol{r}) |V_0| e^{i\phi}
\psi_{2\alpha}(\boldsymbol{r}) + \text{H. c.} \right],
\end{align}
is eliminated by choosing 
$ \psi_{1\alpha} \to \psi_{1\alpha}e^{i\phi/2}$ and 
$ \psi_{2\alpha} \to \psi_{2\alpha}e^{-i\phi/2}$.
The phase $\phi$ does not affect the physics in the normal state. 
When we consider more than one superconducting order parameter, however, 
such a gauge transformation affects the relative phase difference between the order parameters.
This is why we keep the phase degree of freedom of the hybridization potential.
Actually, the phase
plays also an important role in the gap equations as discussed later. 

The magnetic property of Cooper pairs can be discussed 
using the relationship between the electric current and the vector potential 
$\boldsymbol{j}= - \frac{n_e e^2}{mc} Q \boldsymbol{A}$.
Within the linear response regime~\cite{agd,fetter}, the dimensionless pair density $Q$ for a uniform magnetic field 
is calculated to be
\begin{align}
Q=&\frac{1}{2} T\sum_{\omega_n} \int d\xi \, \textrm{Tr}
\left\{ \hat{\mathcal{G}} \hat{\mathcal{G}} + \hat{\mathcal{F}}\underline{\hat{\mathcal{F}}} - 
\hat{\mathcal{G}}_N \hat{\mathcal{G}}_N
\right\}_{(\xi,i\omega_n)},
\end{align}
for the simple two-band model in Eq.~(\ref{smodel}), where $\mathcal{G}_N$ is the Green function in the normal 
state and $\mathrm{Tr}$ means the summation over band degree of freedom. 
When we decompose the anomalous Green functions into
\begin{align}
\hat{\mathcal{F}}=&\sum_{\nu=0}^{3} f_\nu \hat{\rho}_\nu,\quad
\underline{\hat{\mathcal{F}}}=\sum_{\nu=0}^{3} \underline{f}_\nu \hat{\rho}_\nu,
\end{align}
the contribution of the $\nu$ th component of the anomalous Green function to the pair density becomes
\begin{align}
Q_{f}(\nu)=T\sum_{\omega_n} \int d\xi \, f_\nu \underline{f}_\nu. \label{qfnu}
\end{align}
The component of $f_\nu$ is diamagnetic as usual for $Q_{f}(\nu)> 0$.
On the other hand, $f_{\nu}$ is paramagnetic for $Q_{f}(\nu)<0$.
As shown in Eq.~(\ref{f_tra}) and the relation in Eq.~(\ref{ff1}), 
$Q_{f}(2)$ is negative. Therefore the odd-frequency pairing correlation is paramagnetic.
It is easy to confirm that all the even-frequency pairing correlations are diamagnetic. 
The pair density for the simple model in Eq.~(\ref{smodel}) is
represented as
\begin{align}
Q=&T\sum_{\omega_n} \int d\xi
\left[ \frac{X_1-\omega_n^2}{2Z_1} - \omega_n^2 \frac{Y_1}{Z_1^2} \right].\label{qsimple}
\end{align}
The total magnetic response of all the components must be diamagnetic 
to realize stable superconducting states, (i.e., $Q>0$). 
Generally speaking, the pair density $Q$ has more complicated expression than Eq.~(\ref{qsimple})
in the presence of the band asymmetry. Details are shown in Appendix C.
It is possible to discuss the type of magnetic response in general cases by 
checking if $f_j$ contributes positively or negatively to $I_{\mathcal{F}}$ 
in Eq.~(\ref{if_general}). We note that the expression of the response function 
in Eq.~(\ref{k_general}) can be applied to highly asymmetric two-band superconductors.

The self-consistent equations for the intraband pair potentials are represented by
\begin{align}
\Delta_1=&g_1 T\sum_{\omega_n} \frac{1}{V_\mathrm{vol}} \sum_{\boldsymbol{k}} 
\frac{1}{2Z_1} \left[ (X_1-|V_0|^2/2-K) \Delta_1 \right.\nonumber\\
&\left.+\left\{ (v_1^2-v_2^2)/2 + iv_1v_2 \right\}\Delta_2\right],\\
\Delta_2=&g_2 T\sum_{\omega_n} \frac{1}{V_\mathrm{vol}}\sum_{\boldsymbol{k}} 
\frac{1}{2Z_1} \left[ (X_1-|V_0|^2/2+K) \Delta_2 \right.\nonumber\\
& \left.+ \left\{ (v_1^2-v_2^2)/2- iv_1v_2 \right\}\Delta_1\right],\label{delta2}
\end{align}
where $g_\lambda$ is the pairing interaction at the $\lambda$ th band.
When we assume $g_1>g_2$ in the simple model in Eq.~(\ref{smodel}), for instance, 
the transition temperature $T_{c_1}$ for $\Delta_1$ 
is larger than the transition temperature $T_{c_2}$ for $\Delta_2$ in the absence of hybridization. 
Eq.~(\ref{delta2}) indicates that hybridization induces $\Delta_2$ even for $ T_{c_2}< T <T_{c_1}$. 
The relative phase between the two pair potentials is determined by the phase of the hybridization potential.
For $V_0=v_1$ and $v_2=0$, the sign of $\Delta_2$ should be the same as that of $\Delta_1$, which leads to
the decrease of $\Delta_-$. As a result, the amplitude of the odd-frequency component in Eq.(\ref{f_tra}) 
decreases.  
On the other hand for $V_0=iv_2$ and $v_1=0$, the sign of $\Delta_2$ should be opposite to that of $\Delta_1$. 
Therefore $\Delta_+$ decreases, which results in the suppression of the odd-frequency component.
The self-consistent equation minimizes the odd-frequency pairing correlations 
and maximize the condensation energy automatically.
In more general cases for $v_1\neq 0$ and $v_2 \neq 0$, there is a relative phase difference 
between the two pair potentials.

\subsection{Hybridization due to spin-orbit interaction without spin flipping}
Next we consider a situation where $V_0=0$ and $V_3(\boldsymbol{k})\neq 0$.
The spin-orbit interaction hybridizes the two conduction bands.
The Hamiltonian is represented by Eq.~(\ref{h1}) with $V_0=0$.
Here we assume that $V_3(\boldsymbol{k})$ is a real value for simplicity.
The Green functions are represented by Eqs.~(\ref{f_tra}) and (\ref{g1}) with 
$v_1 \to s V_3(\boldsymbol{k})$ and $v_2=0$. 
The symmetry of the intraband pairing correlations remain unchanged from 
the ESEE class.
The interband pairing correlations are calculated as
\begin{align}
&\mathcal{F}_{12,\uparrow\downarrow}(\boldsymbol{k},i\omega_n)=\mathcal{F}_{12,\downarrow\uparrow}(\boldsymbol{k},i\omega_n)
= P_{2e} - P_{2o},\\
&\mathcal{F}_{21,\uparrow\downarrow}(\boldsymbol{k},i\omega_n)=\mathcal{F}_{21,\downarrow\uparrow}(\boldsymbol{k},i\omega_n)
= P_{2e} + P_{2o},\\
&P_{2e}=\frac{1}{2Z_1}\left[ V_3(\xi_+\Delta_+-\xi_-\Delta_-)\right],\\
&P_{2o}=\frac{1}{2Z_1} \left[i\omega_n V_3 \Delta_- \right].
\end{align}
We find that 
\begin{align}
\mathcal{F}_{12,\uparrow\downarrow}+\mathcal{F}_{12,\downarrow\uparrow}
+\mathcal{F}_{21,\uparrow\downarrow}+\mathcal{F}_{21,\downarrow\uparrow} = 4P_{2e},
\end{align}
belongs to the ETOE class and
\begin{align}
\mathcal{F}_{12,\uparrow\downarrow}+\mathcal{F}_{12,\downarrow\uparrow}
-\mathcal{F}_{21,\uparrow\downarrow}-\mathcal{F}_{21,\downarrow\uparrow} = -4P_{2o},
\end{align}
belongs to the OTOO class
 because $V_3$ is an odd-momentum-parity function.
The hybridization caused by the spin-orbit interaction generates two types of 
spin-triplet Cooper pairs as summarized in the bottom two columns of Table~\ref{table2}.
We also conclude that the odd-frequency pairs generated by the hybridization are paramagnetic
in the sense that $Q_f(2)<0$ in Eq.~(\ref{qfnu}).

\subsection{Hybridization due to spin-orbit interaction with spin flipping}
In the end of this section, we consider a situation where $V_0=V_3=0$, 
$V_1(\boldsymbol{k})\neq 0$ and $V_2(\boldsymbol{k})\neq 0$.
The spin-orbit interaction hybridizes the two conduction bands with opposite spin.
The reduced Hamiltonian is represented by
\begin{align}
H_2(\boldsymbol{k})=&\left[
\begin{array}{cccc}
\xi_1 & U^\ast & s\Delta_1 & 0\\
U & \xi_2 & 0 & -s\Delta_2 \\
s\Delta_1^\ast & 0& -\xi_1 & U^\ast \\
0& -s\Delta_2^\ast & U & -\xi_2 
\end{array}
\right], \label{h1_2}\\
U=& V_1(\boldsymbol{k})+is V_2(\boldsymbol{k}).
\end{align}
In the first particle-hole space ($\mathcal{N}_1$), the spin of an electron (a hole) in  
band '1' is $\uparrow$ ($\downarrow$) and
the spin of an electron (a hole) in  
band '2' is $\downarrow$ ($\uparrow$).
In the second particle-hole space ($\mathcal{N}_2$), the
spin of a quasiparticle is opposite to those in $\mathcal{N}_1$.
The sign factor $s$ is 1 and $-1$ in $\mathcal{N}_1$ and $\mathcal{N}_2$, respectively.
The phase of hybridization potential does not affect the physics because it
can be eliminated by a gauge transformation 
for each spin degree of freedom. 
In Eq.~(\ref{h1_2}), therefore, we assume that $V_1$ and $V_2$ 
are real potential.  
The anomalous Green function results in
\begin{align}
\hat{\mathcal{F}}_2&(\boldsymbol{k},i\omega_n)
=\frac{s}{2Z_2} 
\left[
\left\{ -X_2\Delta_- +  K\Delta_+ +|V|^2\Delta_- \right\}\hat{\rho}_0 \right. \nonumber\\
&+ \left\{ i\omega_n V_1 \Delta_- + (\xi_+ \Delta_+ - \xi_-\Delta_- ) is V_2 \right\}\hat{\rho}_1  \nonumber\\
&+ \left\{ i\omega_n s V_2 \Delta_- - (\xi_+\Delta_+ - \xi_- \Delta_-)iV_1 \right\}  
\hat{\rho}_2 \nonumber\\
&+ \left.\left\{ - X_2\Delta_+ +K\Delta_-  \right\}\hat{\rho}_3    \right], 
 \label{f1_2}
%
\end{align}
with $Z_2= X^2_2 - Y_2$ and
\begin{align}
X_2 =& \frac{1}{2}[\omega_n^2 + \xi_+^2 + \xi_-^2 + |\Delta_+|^2 + |\Delta_-|^2 + |V|^2],\label{x1_2}\\
Y_2=& K^2 +|V|^2(\xi_+^2 +|\Delta_-|^2), \label{y1_2}
\end{align}
where $K$ and $D_\pm$ are defined in Eq.~(\ref{k_def}) and $|V|^2=V_1^2+V_2^2$.
The normal Green function is presented in Eq.~(\ref{g1_2}) in Appendix B.
The first term of the $f_1$ component and that of the $f_2$ component in Eq.~(\ref{f1_2}) 
represent the odd-frequency pairing correlations.
By using the relation 
 $\underline{\hat{\mathcal{F}}}_2(\boldsymbol{k}, i\omega_n)
=\hat{\mathcal{F}}^\ast_2(-\boldsymbol{k}, i\omega_n)|_{V_2 \to - V_2}$,
we can confirm that the odd-frequency pairing correlations are paramagnetic. 
In the simple model in Eq.~(\ref{smodel}), the expression of the pair density 
in Eq.~(\ref{qsimple}) is valid by applying Eqs.~(\ref{x1_2}) and (\ref{y1_2}). 
The elements of Green function in Eq.~(\ref{f1_2}) correspond to
\begin{align}
\hat{\mathcal{F}}_2^{\mathcal{N}_1}
=\left[\begin{array}{cc} \mathcal{F}_{11,\uparrow\downarrow} & \mathcal{F}_{12,\uparrow\uparrow} \\
\mathcal{F}_{21,\downarrow\downarrow} & \mathcal{F}_{22,\downarrow\uparrow}
\end{array}\right], \;
\hat{\mathcal{F}}_2^{\mathcal{N}_2}
=\left[\begin{array}{cc} \mathcal{F}_{11,\downarrow\uparrow} & \mathcal{F}_{12,\downarrow\downarrow} \\
\mathcal{F}_{21,\uparrow\uparrow} & \mathcal{F}_{22,\uparrow\downarrow}
\end{array}\right],\nonumber
\end{align}
in ${\mathcal{N}_1}$ and $\mathcal{N}_2$, respectively.
To analyze the pairing symmetry, we rewrite the results of equal-spin pairing correlations into 
\begin{align}
&\mathcal{F}_{12,\uparrow\uparrow} + \mathcal{F}_{21,\uparrow\uparrow}
= \frac{-(V_1-iV_2)}{Z_2}(\xi_+\Delta_+ - \xi_- \Delta_-), \label{esp_uu}\\
&\mathcal{F}_{12,\downarrow\downarrow} + \mathcal{F}_{21,\downarrow\downarrow}
= \frac{(V_1+iV_2)}{Z_2}(\xi_+\Delta_+ - \xi_- \Delta_-),\label{esp_dd}\\
&\mathcal{F}_{12,\uparrow\uparrow} - \mathcal{F}_{21,\uparrow\uparrow}
= \frac{1}{Z_2} i\omega_n \Delta_- (V_1-iV_2),\label{esm_uu}\\
&\mathcal{F}_{12,\downarrow\downarrow} - \mathcal{F}_{21,\downarrow\downarrow}
= \frac{-1}{Z_2} i\omega_n \Delta_- (V_1+iV_2).\label{esm_dd}
\end{align}
The correlations in Eqs.~(\ref{esp_uu}) and (\ref{esp_dd}) belong to the even-frequency spin-triplet 
odd-momentum-parity even-band-parity (ETOE) class because $V_1$ and $V_2$ are 
odd-momentum-parity functions.
The correlations in Eqs.~(\ref{esm_uu}) and (\ref{esm_dd}) belong to the odd-frequency spin-triplet 
odd-momentum-parity odd-band-parity (OTOO) class.
The hybridization between the two bands generates 
these pairing correlations. The symmetry classification is shown in 
Table~\ref{table2}.
The spin-singlet pairing correlations that belong to the ESEE class 
are linked to the pair potentials. 
The self-consistent equation for the pair potentials are represented by
\begin{align}
\Delta_1=&g_1 T\sum_{\omega_n} \frac{1}{V_\mathrm{vol}} \sum_{\boldsymbol{k}} \nonumber\\
&\times \frac{1}{2Z_2} \left[ (X_2-|V|^2/2-K) \Delta_1 +|V|^2
\Delta_2/2 \right], \label{scf_21}\\
\Delta_2=&g_2 T\sum_{\omega_n} \frac{1}{V_\mathrm{vol}}\sum_{\boldsymbol{k}} \nonumber\\
&\times\frac{1}{2Z_2} \left[ (X_2-|V|^2/2+K) \Delta_2 +|V|^2\Delta_1/2\right].
\end{align}
These coupled equations 
allow us to understand the role 
of the odd-frequency pairing correlation.
When the two conduction band are identical to each other (i.e., $\xi_-=0$ and $g_1=g_2$), 
$\Delta_1=\Delta_2$ is expected in the self-consistent equations. 
As a result, the odd-frequency pairing correlation vanishes because $\Delta_-=0$ as shown in Eq.~(\ref{f1_2}).
In such a case, we obtain the equation that determines the transition temperature $T_{c_0}$
at $g_1=g_2$,
\begin{align}
1= 2\pi N_0 g_1 T\sum_{\omega_n>0}^{N_{c_0}} \frac{1}{\omega_n}, 
\end{align} 
where $N_0$ is the density of 
states at the Fermi level in each band per spin, $N_{c_0}=\omega_c/(2\pi T_{c_0})$ and $\omega_c$ is the cut-off energy.
The equation is identical to the equation for determining the transition temperature 
in a single-band superconductor. Therefore the hybridization does not affect transition temperature
when $\Delta_1=\Delta_2$ is expected in superconducting states.
On the other hand for $g_1 \gg g_2$, $\Delta_1 \gg \Delta_2$ is expected in superconducting states.
By neglecting $\Delta_2$ in Eq.~(\ref{scf_21}), we find that
\begin{align}
1= 2\pi N_0 g_1 T\sum_{\omega_n>0}^{N_c} \left[ \frac{1}{\omega_n}
- \frac{|V|^2/2}{\omega_n(\omega_n^2+|V|^2)} \right],\label{tc_odd}
\end{align} 
determines the transition temperature $T_c$ with $N_{c}=\omega_c/(2\pi T_{c})$. 
Because of the second term in Eq.~(\ref{tc_odd}), $N_c > N_{c_0}$ is necessary 
for Eq.~(\ref{tc_odd}) to have a solution. The condition is 
identical to $T_c < T_{c_0}$.
In this way, we conclude that the hybridization reduces the transition temperature 
when $\Delta_1\neq \Delta_2$ is expected in superconducting states.
In such states, the induced odd-frequency pairs 
are unstable because of their paramagnetic instability. 
The expression of the pair density in Eq.~(\ref{qsimple}) with Eq.~(\ref{smodel}) 
is also valid for this case when we replace $X_1$ and $Y_1$ by corresponding values
Eqs.~(\ref{x1_2}) and (\ref{y1_2}).
A stable superconducting state is possible as long as $Q$ is positive. 
If we could control the fraction of odd-frequency pairs in real materials, 
it would be possible to detect them from the variation of $T_c$.

\section{Interband pairing order}
In this section, we consider the superconducting order parameter that derived from the attractive interaction
between two electrons in different conduction bands. 
The total Hamiltonian is given by Eq.~(\ref{bcs88}) with $\Delta_1=0$ and $\Delta_2=0$.
Even if we focus on equal-time $s$-wave pair potential, the band degree of freedom 
makes both the spin-triplet and the spin-singlet pair potentials possible.
Here we assume that the attractive interaction acts on 
two electrons with different spins in different conduction bands. 
Actually the interband spin-triplet order parameters~\cite{fu} have been discussed in Cu-doped Bi$_2$Se$_3$. 

\subsection{Spin-singlet pairing order}
\begin{table*}[t]
\begin{ruledtabular}
\begin{tabular}{lccccc}
\null & Frequency  & Spin &  Momentum parity & Band parity & Magnetic response \\
\hline
Pair potential &  Even & Singlet & Even  & Even (inter) & Diamagnetic \\
\hline
Induced by $V_0$ &  Even & Singlet & Even & Even(intra)   & Diamagnetic \\
Induced by $V_1$, $V_2$, $V_3$ &  Even & Triplet & Odd & Even(intra)   & Diamagnetic \\
Induced by $\xi_-$ &   Odd & Singlet & Even &  Odd  & Paramagnetic \\
\end{tabular}
\end{ruledtabular}
\caption[b]{
The symmetry classification of Cooper pairs with an 
equal-time spin-singlet $s$-wave even-band-parity (interband) pair potential.
}
\label{table3}
\end{table*}
We first consider the situation where the hybridization preserves spin by
setting $V_1=V_2=0$. The hamiltonian is described by
\begin{align}
\check{H}_3(\boldsymbol{k})=&\left[
\begin{array}{cccc}
\xi_1 & W & 0&   s\Delta \\
W^\ast & \xi_2 & s\Delta & 0 \\
0&  s \Delta^\ast &  -\xi_1 & -W^\ast \\
 s \Delta^\ast & 0 & -W & -\xi_2 
\end{array}
\right], \label{h3}\\
W=&V_0 + sV_3(\boldsymbol{k}),
\end{align}
for the spin-singlet pair potential, (i.e., $s_{\mathrm{spin}}=-1$ in Eq.~(\ref{bcs88})).
Hereafter we remove '12' from the subscript of the interband pair potential for simplicity.
In the particle-hole space $\mathcal{N}_1$, the spin of an electron is $\uparrow$ in the two 
 conduction bands, 
whereas that of a hole is $\downarrow$.
The spin direction is opposite in $\mathcal{N}_2$.
At $V_3=0$, the anomalous Green functions is calculated as
\begin{align}
\hat{\mathcal{F}}_3&(\boldsymbol{k},i\omega_n)=
\frac{s \Delta}{2Z_3}\left[ (v_1\xi_+ - i v_2 \xi_-) \hat{\rho}_0 
+ (-X_3+\xi_-^2) \hat{\rho}_1 \right.\nonumber\\
& \left.- \omega_n \xi_- \hat{\rho}_2
+(-v_1 \xi_- +i v_2  \xi_+)  \hat{\rho}_3 
 \right],\label{f3}
\end{align}
with $Z_3=X^2_3-Y_3$ and
\begin{align}
X_3=&\frac{1}{2}[\omega_n^2 + \xi_+^2 + \xi_-^2 + |V_0|^2 + |\Delta|^2], \label{x3}\\
Y_3=& \xi_+^2 \xi_-^2 + \xi_+^2 |V_0|^2 + \xi_-^2 |\Delta|^2. \label{y3}
\end{align}
The normal part is given in Eq.~(\ref{g3}) in Appendix B.
The phase of the hybridization potential does not play any role in this case.
Actually, $Z_3$ depends only on $|V_0|$ as shown in Eqs.~(\ref{x3}) and (\ref{y3}).
The elements of the anomalous Green function in Eq.~(\ref{f3}) correspond to
\begin{align}
\hat{\mathcal{F}}_3^{\mathcal{N}_1}
=\left[\begin{array}{cc} \mathcal{F}_{11,\uparrow\downarrow} & \mathcal{F}_{12,\uparrow\downarrow} \\
\mathcal{F}_{21,\uparrow\downarrow} & \mathcal{F}_{22,\uparrow\downarrow}
\end{array}\right], \,
\hat{\mathcal{F}}_3^{\mathcal{N}_2}
=\left[\begin{array}{cc} \mathcal{F}_{11,\downarrow\uparrow} & \mathcal{F}_{12,\downarrow\uparrow} \\
\mathcal{F}_{21,\downarrow\uparrow} & \mathcal{F}_{22,\downarrow\uparrow}
\end{array}\right], \label{f3_structure}
\end{align}
in ${\mathcal{N}_1}$ and $\mathcal{N}_2$, respectively.

The intraband pairing correlations become
\begin{align}
\mathcal{F}_{11,\uparrow\downarrow} -\mathcal{F}_{11,\downarrow\uparrow}
=& \frac{\Delta}{Z_3}(\xi_+-\xi_-)V_0,\\
\mathcal{F}_{22,\uparrow\downarrow} -\mathcal{F}_{22,\downarrow\uparrow}
=& \frac{\Delta}{Z_3}(\xi_++ \xi_-)V_0^\ast.
\end{align}
They are generated by the hybridization and belong to the ESEE class.
The band asymmetry generates 
the interband pairing correlation
\begin{align}
\mathcal{F}_{12,\uparrow\downarrow} -\mathcal{F}_{12,\downarrow\uparrow}
-\mathcal{F}_{21,\uparrow\downarrow} +\mathcal{F}_{21,\downarrow\uparrow}
=& \frac{2\Delta}{Z_3}i\omega_n \xi_-,
\end{align}
which belongs to the OSEO class.
The odd-frequency pairing correlation is paramagnetic because 
$\underline{\mathcal{F}}(\boldsymbol{k},i\omega_n) = \mathcal{F}^\ast(\boldsymbol{k},i\omega_n)$.
The pair potential is linked to the pairing correlation
\begin{align}
\mathcal{F}_{12,\uparrow\downarrow} -\mathcal{F}_{12,\downarrow\uparrow}
+\mathcal{F}_{21,\uparrow\downarrow} -\mathcal{F}_{21,\downarrow\uparrow}
=& \frac{2\Delta}{Z_3}(-X_3 + \xi_-^2).
\end{align}
The self-consistent equation becomes
\begin{align}
1=g T\sum_{\omega_n} \frac{1}{V_\mathrm{vol}} \sum_{\boldsymbol{k}} 
\frac{X_3-\xi_-^2}{2Z_3}.\label{scf3}
\end{align}

When we consider spin-orbit hybridization $V_3$ at $V_0=0$,
the Green functions are given by Eqs.~(\ref{g3}) and (\ref{f3}) 
with $v_1 \to sV_3$ and $v_2 \to 0$.
The intraband pairing correlations are
\begin{align}
\mathcal{F}_{11,\uparrow\downarrow}+\mathcal{F}_{11,\downarrow\uparrow}
=& \frac{\Delta}{Z_3}(\xi_+-\xi_-)V_3,\\
\mathcal{F}_{22,\uparrow\downarrow}+\mathcal{F}_{22,\downarrow\uparrow}
=& \frac{\Delta}{Z_3}(\xi_++\xi_-)V_3.
\end{align}
The hybridization $V_3$ generates pairing correlations that belong to the ETOE class. 
The results of the symmetry classification are summarized in Table.~\ref{table3}.

Next we consider effects of spin-flip hybridization on the pairing correlations.
At $V_0=V_3=0$, the Hamiltonian can be reduced to
\begin{align}
\check{H}_4(\boldsymbol{k})=&\left[
\begin{array}{cccc}
\xi_1 & U^\ast & 0&   s\Delta \\
U & \xi_2 & - s\Delta & 0 \\
0& - s \Delta^\ast &  -\xi_1 & U \\
 s\Delta^\ast & 0 & U^\ast & -\xi_2 
\end{array}
\right], \label{h4}\\
U=& V_1(\boldsymbol{k})+is V_2(\boldsymbol{k}).
\end{align}
In the first particle-hole space ($\mathcal{N}_1$), the spin of a quasiparticle in 
band '1' is $\uparrow$ and that in band '2' is $\downarrow$
for both an electron and a hole.
In the second particle-hole space ($\mathcal{N}_2$), 
the spins of a quasiparticle are opposite to those in $\mathcal{N}_1$.
The Green function is calculated as follows
\begin{align}
\hat{\mathcal{F}}_4&(\boldsymbol{k},i\omega_n)=
\frac{s \Delta}{2Z_4}\left[
(V_1\xi_- +is V_2\xi_+) \hat{\rho}_0   +i \omega_n \xi_- \hat{\rho}_1 \right.\nonumber\\ 
&\left. -i (X_4-\xi_-^2) \hat{\rho}_2 
-(V_1\xi_+ + i s V_2\xi_-) \hat{\rho}_3 
 \right],\label{f4}
\end{align}
with $Z_4=X^2_4-Y_4$, $|V|^2\equiv V_1^2+V_2^2$ and
\begin{align}
X_4=&\frac{1}{2}[\omega_n^2 + \xi_+^2 + \xi_-^2 + |V|^2 + |\Delta|^2], \label{x4}\\
Y_4=& \xi_+^2 \xi_-^2 + \xi_+^2 |V|^2 + \xi_-^2 |\Delta|^2. \label{y4}
\end{align}
The normal part is shown in Eq.~(\ref{g4}) in Appendix B.
The elements of the anomalous Green function in Eq.~(\ref{f4}) are represented by
\begin{align}
\hat{\mathcal{F}}_4^{\mathcal{N}_1}
=\left[\begin{array}{cc} \mathcal{F}_{11,\uparrow\uparrow} & \mathcal{F}_{12,\uparrow\downarrow} \\
\mathcal{F}_{21,\downarrow\uparrow} & \mathcal{F}_{22,\downarrow\downarrow}
\end{array}\right], \;
\hat{\mathcal{F}}_4^{\mathcal{N}_2}
=\left[\begin{array}{cc} \mathcal{F}_{11,\downarrow\downarrow} & \mathcal{F}_{12,\downarrow\uparrow} \\
\mathcal{F}_{21,\uparrow\downarrow} & \mathcal{F}_{22,\uparrow\uparrow}
\end{array}\right], \nonumber
\end{align}
in $\mathcal{N}_1$ and $\mathcal{N}_2$, respectively.
It is easy to confirm that 
the hybridizations $V_1$ and $V_2$ generate the equal-spin intraband pairing correlations
\begin{align}
F_{11,\uparrow\uparrow}=&\frac{\Delta}{2Z_4}(\xi_+ - \xi_-)(-V_1+iV_2),\label{f4_11u}\\
F_{11,\downarrow\downarrow} =&\frac{\Delta}{2Z_4}(\xi_+ - \xi_-)(V_1+iV_2),\label{f4_11d}\\
F_{22,\uparrow\uparrow} =& \frac{\Delta}{2Z_4}(\xi_+ + \xi_-)(-V_1+iV_2)\label{f4_22u},\\
F_{22,\downarrow\downarrow}=& \frac{\Delta}{2Z_4}(\xi_+ + \xi_-)(V_1+iV_2).\label{f4_22d}
\end{align}
They belong to the ETOE class.
The band asymmetry generates the pairing correlation
\begin{align}
F_{12,\uparrow\downarrow} -F_{12,\downarrow\uparrow}- F_{21,\uparrow\downarrow} +F_{21,\downarrow\uparrow} 
=\frac{2\Delta}{Z_4} i \omega_n \xi_-,
\end{align}
which belongs to the OSEO class.
Since $\underline{\mathcal{F}}_4(\boldsymbol{k},i\omega_n) = -\mathcal{F}_4^\ast(-\boldsymbol{k},i\omega_n)$, 
the odd-frequency pairing correlation is paramagnetic.
The results of the symmetry classification are shown in Table.\ref{table3}.
The self-consistent equation is described by Eq.~(\ref{scf3}) with $X_3\to X_4$ and $Z_3 \to Z_4$.

\subsection{Spin-triplet pairing order}

\begin{table*}[t]
\begin{center}
\begin{ruledtabular}
\begin{tabular}{lccccc}
\null & Frequency  & Spin & Momentum parity & Band parity & Magnetic response \\
\hline
Pair potential &  Even & Triplet & Even & Odd  & Diamagnetic \\
\hline
Induced by $V_0$ &  Odd & Triplet & Even & Even (intra)  & Paramagnetic \\
Induced by $\xi_-$ &   Odd & Triplet & Even & Even (inter) & Paramagnetic \\
Induced by $V_1, V_2, V_3$ &  Even & Triplet & Odd & Even (intra) & Diamagnetic 
\end{tabular}
\end{ruledtabular}
\end{center}
\caption[b]{
The symmetry classification of Cooper pairs with an equal-time spin-triplet $s$-wave odd-band-parity 
pair potential.
}
\label{table4}
\end{table*}

Finally we consider the spin-triplet odd-band-parity pair potential.
When the hybridization preserves spin (i.e., $V_1=V_2=0$), the reduced Hamiltonian is given by 
\begin{align}
\check{H}_5(\boldsymbol{k})=&\left[
\begin{array}{cccc}
\xi_1 & W & 0&   \Delta \\
W^\ast & \xi_2 & -\Delta & 0 \\
0&  - \Delta^\ast &  -\xi_1 & -W^\ast \\
  \Delta^\ast & 0 & -W & -\xi_2 
\end{array}
\right], \label{h5}
\end{align}
with $W=V_0+sV_3(\boldsymbol{k})$.
At $V_3=0$, the Green function for $V_0\neq 0$ becomes
\begin{align}
\hat{\mathcal{F}}_5&(\boldsymbol{k},i\omega_n)=
\frac{\Delta}{2Z_5}\left[ \omega_n v_2
 \hat{\rho}_0 + i\omega_n \xi_- \hat{\rho}_1 \right.\nonumber\\
& \left. -i(X_5 -|V_0|^2 - \xi_-^2) \hat{\rho}_2
-i \omega_n v_1  \hat{\rho}_3 
 \right],\label{f5}
\end{align}
with $Z_5=X^2_5-Y_5$ and
\begin{align}
X_5=&\frac{1}{2}[\omega_n^2 + \xi_+^2 + \xi_-^2 + |V_0|^2 + |\Delta|^2], \label{x5}\\
Y_5=& (\xi_+^2 +|\Delta|^2)(\xi_-^2 + |V_0|^2). \label{y5}
\end{align}
The elements of the anomalous Green function in Eq.~(\ref{f5}) correspond to 
Eq.~(\ref{f3_structure}).
In this case, both the hybridization and the band asymmetry generate the odd-frequency 
pairing correlations.
The hybridization induces the correlations
\begin{align}
\mathcal{F}_{11,\uparrow\downarrow}+\mathcal{F}_{11,\downarrow\uparrow} =&-\frac{\Delta}{Z_5}i\omega_n V_0,\\
\mathcal{F}_{22,\uparrow\downarrow}+\mathcal{F}_{22,\downarrow\uparrow} =&\frac{\Delta}{Z_5}i\omega_n V_0^\ast.
\end{align}
The band asymmetry induces 
\begin{align}
\mathcal{F}_{12,\uparrow\downarrow}+\mathcal{F}_{12,\downarrow\uparrow} 
+\mathcal{F}_{21,\uparrow\downarrow}+\mathcal{F}_{21,\downarrow\uparrow} 
=&\frac{2\Delta}{Z_5}i\omega_n \xi_-.
\end{align}
All of them belong to the OTEE class.
Since $\underline{\mathcal{F}}_5(\boldsymbol{k},i\omega_n) = -\mathcal{F}_5^\ast(\boldsymbol{k},i\omega_n)$,
all of the odd-frequency pairing correlations are paramagnetic.

The Green function for $V_0=0$ and $V_3\neq 0$ are presented in 
Eq.~(\ref{g5}) and (\ref{f5}) with $v_1 \to sV_3$ and $v_2 \to 0$.
The spin-orbit hybridization generates the correlations
\begin{align}
\mathcal{F}_{11,\uparrow\downarrow}-\mathcal{F}_{11,\downarrow\uparrow} =&-\frac{\Delta}{Z_5}i\omega_n V_3,\\
\mathcal{F}_{22,\uparrow\downarrow}-\mathcal{F}_{22,\downarrow\uparrow} =&\frac{\Delta}{Z_5}i\omega_n V_3.
\end{align}
They belong to the OSOE class.

Finally we discuss the effects of spin-flip hybridization $V_1$ and $V_2$ on the pairing correlations.
The Hamiltonian is given by Eq.~(\ref{h4}) with $s\Delta \to \Delta$.
The Green function is presented in Eq.~(\ref{g4}) and (\ref{f4}) with $s\Delta\to \Delta$.
The spin-flip hybridization generate the pairing correlations that belong to the ETOE class.
The correlations $F_{11,\uparrow\uparrow}$ and $F_{22,\downarrow\downarrow}$ are equal to 
Eqs.~(\ref{f4_11u}) and (\ref{f4_22d}), respectively.
The remaining components $F_{11,\downarrow\downarrow}$ and $F_{22,\uparrow\uparrow}$ change their 
signs from Eqs.~(\ref{f4_11d}) and (\ref{f4_22u}), respectively.
The band asymmetry induces 
\begin{align}
F_{12,\uparrow\downarrow}+F_{12,\downarrow\uparrow}+F_{21,\uparrow\downarrow}+F_{21,\downarrow\uparrow}
=\frac{2\Delta}{Z_4} i \omega_n \xi_-,
\end{align}
which belongs to the OTEE class and has a paramagnetic property because of 
$\underline{\mathcal{F}}_4(\boldsymbol{k},i\omega_n) = -\mathcal{F}_4^\ast(-\boldsymbol{k},i\omega_n)$.
The symmetry classification results for the spin-triplet order parameter are summarized 
in Table.~\ref{table4}. The spin-orbit hybridization generates the equal-spin pairing correlation.


\section{Conclusion}
We have studied the magnetic response of the odd-frequency pairing correlations appearing 
in two-band superconductors by using analytical expressions of the anomalous Green function
for the Gor'kov equation. 
We confine ourselves to the equal-time $s$-wave pair potential and introduce 
two types of hybridization potentials between the two conduction bands.
One is a spin-independent hybridization potential.
The other is a spin-dependent hybridization potential derived from the spin-orbit interaction.
The hybridization potentials and the asymmetry in the two conduction bands 
generate the odd-frequency pairing correlations. We conclude that odd-frequency pairs 
always exhibit a paramagnetic response to external magnetic fields irrespective of 
their generation process. 
By analyzing the self-consistent equation for the order parameter, 
we found that the odd-frequency pairing correlations reduce 
the superconducting transition temperature because of their paramagnetic instability.
The hybridization potential also generates 
even-frequency Cooper pairs, which belong to a different symmetry class from that of the pair potential.
The appearance of the subdominant pairing correlation can be understood as 
a deformation of the ground state by the hybridization and/or the band asymmetry~\cite{diaodd}.
Thus rich symmetry variety of Cooper pairs in a uniform ground state 
is an essential feature of two-band superconductors.

\begin{acknowledgments}
The authors are grateful to A. V. Balatsky, Y. Tanaka, Y. V. Fominov, and A. A. Golubov for useful discussions.
This work was supported by ``Topological Materials Science'' (No.~15H05852) 
and KAKENHI (Nos.~26287069 and 15H03525) from
the Ministry of Education,
Culture, Sports, Science and Technology (MEXT) of Japan
and by the Ministry of Education and Science of the Russian Federation
(Grant No.~14Y.26.31.0007).
\end{acknowledgments}

\appendix
\section{Paramagnetic property of odd-frequency Cooper pairs}
The superconducting state is described well by a phenomenological many-body wave function
\begin{align}
\Psi(\boldsymbol{r})=\sqrt{n_s}e^{i\varphi(\boldsymbol{r})},\label{wf}
\end{align}
where $n_s$ is the uniform pair density and $\varphi$ is the macroscopic phase.
The phase coherence is an essential property of superconductivity, which is described by
$\nabla \varphi=0$ in the uniform ground state.
The current density is defined by
\begin{align}
\boldsymbol{j}= \frac{e\hbar}{2mi}
\left( \Psi^\ast \nabla \Psi - \nabla \Psi^\ast \Psi \right) -\frac{e^2|\Psi|^2}{mc} \boldsymbol{A},
\end{align}
under an external magnetic field $\boldsymbol{H}=\nabla \times \boldsymbol{A}$.
By applying the rigid phase coherence $\nabla \varphi\approx 0$, 
we obtain the so called London equation,
\begin{align}
\boldsymbol{j}=  -\frac{ n_s e^2}{mc} \boldsymbol{A}.\label{london}
\end{align}
By substituting Eq.~(\ref{london}) into the Maxwell equation $\nabla\times \boldsymbol{H}=(4\pi/c) \boldsymbol{j}$,
we obtain 
\begin{align}
&\nabla^2 \boldsymbol{A}-\frac{1}{\lambda_L^2} \boldsymbol{A}=0,\label{maxwell}
\end{align}
where $\lambda_L= (4\pi n_s e^2/mc^2)^{-2}$ is the London penetration length.
Eq.~(\ref{maxwell}) indicates the exponential decay of a magnetic field into a superconductor.
Thus all superconductors are diamagnetic.
The negative sign on the right-hand-side of Eq.~(\ref{london}) is essential in this argument. 
Indeed, the charge square $e^2$, the electron mass $m$, the speed of light $c$, and the pair density $n_s$
are all positive values. The superconducting condensate decreases its free energy by keeping the rigid 
phase coherence, which results in the diamagnetism of a superconductor.
Eq.~(\ref{wf}) means nothing other than the wave function of Cooper pairs which belong 
to the equal-time (even-frequency) spin-singlet $s$-wave class.
Thus even-frequency Cooper pairs in a uniform superconductor are diamagnetic.
 
 In a microscopic theory, the electric current is connected with the vector potential by 
a response function $K$ as shown in Appendix C. The Cooper pairs are described by the 
anomalous Green function instead of the many-body wave function in Eq.~(\ref{wf}). 
In an inhomogeneous superconductor, odd-frequency Cooper pairs are generated by the spatial gradient 
of pair potential. For instance,
 odd-frequency pairs appear at a surface of 
spin-singlet $d$-wave or spin-triplet $p$-wave superconductor as a subdominant pairing correlation.
Ref.~\onlinecite{suzuki1} demonstrated a paramagnetic response of odd-frequency pairs to an external 
magnetic field at surfaces of such unconventional superconductors. 
At the London equation in Eq.~(\ref{london}), the paramagnetic property is represented by 
\textsl{negative} pair density $n_s<0$, which immediately results in the penetration of magnetic field 
into a superconductor in Eq.~(\ref{maxwell}). The paramagnetic Cooper pairs attract a magnetic field and 
break the phase coherence. As a consequence, the existence of odd-frequency pairs increases the free energy as 
demonstrated in Ref.~\onlinecite{suzuki2}. Therefore odd-frequency pairs 
are thermodynamically unstable. In single-band superconductors, the odd-frequency pairs always appear 
locally due to their instability.

To our knowledge, the odd-frequency pairing correlations derived from the Bogoliubov-de Gennes Hamiltonian
are always paramagnetic and thermodynamically unstable.
Thus the presence of uniform odd-frequency pairs in two-band superconductors is a surprising conclusion.
To realize stable and diamagnetic two-band superconducting states, the total pair 
density (corresponding to $Q$ in the text) must be positive.

Finally a relation to Majorana physics~\cite{lutchyn2010,oreg2010} may be worth mentioning.
We have shown that Majorana fermions in a topologically nontrivial superconducting nanowire 
always accompany odd-frequency pairs~\cite{asano13}.
To our knowledge, subgap quasiparticles in single-band superconductors look like 
odd-frequency Cooper pairs.

\section{Normal Green function}
The normal Green functions are summarized in this appendix.

The results for $H_1$ is given by
\begin{align}
\hat{\mathcal{G}}_1&(\boldsymbol{k},i\omega_n)
=\frac{1}{2Z_1} 
\left[
\left\{ A_1 + |V_0|^2 \xi_+  +  \xi_- K \right\} \hat{\rho}_0 \right.\nonumber\\
&- \left\{ (B_1 - |\Delta_-|^2 )v_1 +iD_- v_2 \right\} \hat{\rho}_1  \nonumber \\
&+ \left\{ (B_1 - |\Delta_+|^2 )v_2 + i D_- v_1 \right\}\hat{\rho}_2 \nonumber\\
&\left.
+ \left\{ -X_1 \xi_- + ( i\omega_n + \xi_+ ) K \right\} \hat{\rho}_3 \right],\label{g1}\\
A_1=& -(i\omega_n + \xi_+) X_1, \; B_1= X_1 - (i\omega_n+\xi_+) \xi_+,
\end{align}
at $V_3=0$.
The two normal Green functions satisfy the relations
$\underline{\hat{\mathcal{G}}}_1(\boldsymbol{k},i\omega_n)
=- \hat{\mathcal{G}}^\ast_1(\boldsymbol{k},i\omega_n)$.

The normal Green fonction for $H_2$ is given by
\begin{align}
\hat{\mathcal{G}}_2&(\boldsymbol{k},i\omega_n)
=\frac{1}{2Z_2} 
\left[
\left\{ A_2 + |V|^2 \xi_+  +  \xi_- K \right\} \hat{\rho}_0 \right.\nonumber\\
&- \left\{ ( B_2 - |\Delta_-|^2 )V_1 +iD_- s V_2 \right\} \hat{\rho}_1  \nonumber \\
&- \left\{ (B_2 - |\Delta_-|^2 )s V_2 - iD_- V_1 \right\}\hat{\rho}_2 \nonumber\\
&\left.
+ \left\{ -X_2 \xi_- + ( i\omega_n + \xi_+ ) K \right\} \hat{\rho}_3 \right],\label{g1_2}\\
A_2=& -(i\omega_n + \xi_+) X_2, \; B_2= X_2 - (i\omega_n+\xi_+) \xi_+,
\end{align}
and satisfies the relation
$\underline{\hat{\mathcal{G}}}_2(\boldsymbol{k}, i\omega_n)
= -\hat{\mathcal{G}}^\ast_2(-\boldsymbol{k}, i\omega_n)|_{V_2 \to - V_2}$ 

For $H_3$, the normal Green function at $V_3=0$ is given by
\begin{align}
\hat{\mathcal{G}}_3&(\boldsymbol{k},i\omega_n)=
\frac{1}{2Z_3}\left[
\left\{ A_3 + \xi_+(\xi_-^2+|V_0|^2) \right\}\hat{\rho}_0 \right.\nonumber\\
&+ B_3 v_1 \hat{\rho}_1 
 -B_3v_2 \hat{\rho}_2
\left.+ (B_3 -|\Delta|^2) \xi_- \hat{\rho}_3 \right],\label{g3}\\ 
A_3=& -(i\omega_n + \xi_+) X_3, \; B_3= X_3 - (i\omega_n+\xi_+) \xi_+.
\end{align}
The two normal Greens function are related by
$\underline{\mathcal{G}}_3(\boldsymbol{k},i\omega_n) = -\mathcal{G}_3^\ast(\boldsymbol{k},i\omega_n)$.

One normal Green function for $H_4$ is calculated as
\begin{align}
\hat{\mathcal{G}}_4&(\boldsymbol{k},i\omega_n)=
\frac{1}{2Z_4}\left[
\left\{ A_4 + \xi_+(\xi_-^2+|V|^2) \right\}\hat{\rho}_0 \right.\nonumber\\
&- B_4 V_1 \hat{\rho}_1
-B_4 s V_2 \hat{\rho}_2 
\left. -(B_4-|\Delta|^2) \xi_- \hat{\rho}_3 \right],\label{g4}\\
A_4=& -(i\omega_n + \xi_+) X_4, \; B_4= X_4 - (i\omega_n+\xi_+) \xi_+.
\end{align}
The other normal Green function can be obtained by using 
$\underline{\hat{\mathcal{G}}}_4\boldsymbol{k},i\omega_n)
=- \hat{\mathcal{G}}^\ast_4(-\boldsymbol{k},i\omega_n)$.

For $H_5$, the normal Green function at $V_3=0$ is represented by 
\begin{align}
\hat{\mathcal{G}}_5&(\boldsymbol{k},i\omega_n)=
\frac{1}{2Z_5}
\left[
\left\{ A_5 + \xi_+(\xi_-^2+|V_0|^2) \right\}\hat{\rho}_0 \right.\nonumber\\
&\left. -(B_5-|\Delta|^2 ) (v_1 \hat{\rho}_1 -v_2 \hat{\rho}_2 +\xi_- \hat{\rho}_3)
\right],\label{g5}\\ 
A_5= -(i&\omega_n + \xi_+) X_5, \; B_5= X_5 - (i\omega_n+\xi_+) \xi_+.
\end{align}
The two normal Green function satisfy
$\underline{\hat{\mathcal{G}}}_5\boldsymbol{k},i\omega_n)
=- \hat{\mathcal{G}}^\ast_5(\boldsymbol{k},i\omega_n)$.

\section{Pair density}
In this paper, we consider a two-band superconductor in type II.
The relation
\begin{align}
j_\mu(\mathrm{x})=- \int d\mathrm{x}^\prime K_{\mu\nu}(\mathrm{x},\mathrm{x}^\prime)
A_\nu(\mathrm{x}^\prime),
\end{align}
with $\mathrm{x}=(\boldsymbol{r},t)$ connects the
electric current and the vector potential.
Within the linear response to vector potential~\cite{agd,fetter}, the electric current is described by
\begin{align}
&j_\mu(\boldsymbol{q},\omega_l)
=-K_{\mu\nu}(\boldsymbol{q},\omega_l)
A_\nu(\boldsymbol{q},\omega_l),\\
&K_{\mu\nu}(\boldsymbol{q},\omega_l)= \frac{e^2}{c}
\left[ \sum_{\lambda=1,2}\sum_{\alpha=\uparrow,\downarrow} \frac{n_{\lambda,\alpha}}{m_\lambda}\delta_{\mu\nu} \right.\nonumber\\
&+ T\sum_{\omega_n} \frac{1}{V_{\mathrm{vol}}} \sum_{\boldsymbol{k}} 
\sum_{\alpha,\alpha'}\sum_{\lambda,\lambda'}
\frac{(\boldsymbol{k} + \boldsymbol{q}/2)_\mu}{m_\lambda}
\frac{(\boldsymbol{k} + \boldsymbol{q}/2)_\nu}{m_{\lambda'}} \nonumber\\
&\times
\left\{ 
\mathcal{G}_{\lambda\alpha, \lambda'\alpha'}(\boldsymbol{k} + \boldsymbol{q}, \omega_n+\omega_l)
\mathcal{G}_{\lambda'\alpha', \lambda\alpha}(\boldsymbol{k}, \omega_n) \right.\nonumber\\
&+\left. \left.
\mathcal{F}_{\lambda\alpha, \lambda'\alpha'}(\boldsymbol{k} + \boldsymbol{q}, \omega_n+\omega_l)
\underline{\mathcal{F}}_{\lambda'\alpha', \lambda\alpha}(\boldsymbol{k}, \omega_n)
\right\} 
\right],
\end{align}
for a two-band superconductor, where $n_{\lambda,\alpha}$ is the electron density with spin $\alpha$ in the band $\lambda$ 
and $\omega_l$ is a bosonic Matsubara frequency.
In the static limit $\omega_l\to 0$ and the uniform magnetic field $\boldsymbol{q}\to 0$, 
the diagonal element describes the Meissner effect.
In such a case, the Meissner kernel becomes
\begin{align}
&K\equiv K_{\mu\mu}(0,0)= \frac{e^2}{c}
\left[ \sum_{\lambda=1,2}\sum_{\alpha=\uparrow,\downarrow} \frac{n_{\lambda,\alpha}}{m_\lambda} \right.\nonumber\\
&+ T\sum_{\omega_n} \frac{1}{V_{\mathrm{vol}}} \sum_{\boldsymbol{k}} 
\sum_{\alpha,\alpha'}\sum_{\lambda,\lambda'}
\frac{ k}{m_\lambda}
\frac{ k}{m_{\lambda'}}\frac{1}{d} \nonumber\\
&\times
\left\{ 
\mathcal{G}_{\lambda\alpha, \lambda'\alpha'}(\boldsymbol{k}, \omega_n)
\mathcal{G}_{\lambda'\alpha', \lambda\alpha}(\boldsymbol{k}, \omega_n) \right.\nonumber\\
&+\left. \left.
\mathcal{F}_{\lambda\alpha, \lambda'\alpha'}(\boldsymbol{k}, \omega_n)
\underline{\mathcal{F}}_{\lambda'\alpha', \lambda\alpha}(\boldsymbol{k}, \omega_n)
\right\} 
\right],
\end{align}
where $d$ is the dimensionality of a superconductor.
When we decompose the Green function into
\begin{align}
\mathcal{G}_{\lambda,\lambda'}=\sum_{j=0}^{3}
g_j \hat{\rho}_j, \quad
\mathcal{F}_{\lambda,\lambda'}=\sum_{j=0}^{3}
f_j \hat{\rho}_j, 
\end{align}
in two-band space, we obtain
\begin{align}
&I_{\mathcal{F}} \equiv \sum_{\lambda,\lambda'}
\frac{ 1}{m_\lambda}
\frac{ 1}{m_{\lambda'}}
\mathcal{F}_{\lambda\alpha, \lambda'\alpha'}(\boldsymbol{k}, \omega_n)
\underline{\mathcal{F}}_{\lambda'\alpha', \lambda\alpha}(\boldsymbol{k}, \omega_n),\\
&=
\left[ \frac{1}{m_1^2}  + \frac{1}{m_2^2}\right]
(f_0 \underline{f}_0 + f_3 \underline{f}_3)+
\frac{2}{m_1m_2}
(f_1 \underline{f}_1 + f_2 \underline{f}_2) \nonumber\\
&+
\left[ \left(\frac{1}{m_1} \right)^2 - \left(\frac{1}{m_2} \right)^2\right]
(f_0 \underline{f}_3 + f_3 \underline{f}_0), \label{if_general} 
\end{align}
where we omit spin indices $\alpha$ and $\alpha'$.
The kernel becomes
\begin{align}
K= &\frac{e^2}{c}
\left[ \sum_{\lambda=1,2}\sum_{\alpha=\uparrow,\downarrow} \frac{n_{\lambda,\alpha}}{m_\lambda} \right.\nonumber\\
&+ T\sum_{\omega_n} \frac{1}{V_{\mathrm{vol}}} \sum_{\boldsymbol{k}} 
\left. \sum_{\alpha,\alpha'}\frac{k^2}{d} \left\{ I_{\mathcal{G}} + I_{\mathcal{F}} \right\}
\right]. \label{k_general}
\end{align}
Eq.~(\ref{k_general}) can be applied to highly asymmetric two-band superconductors.

For a simple model in Eq.~(\ref{smodel}), we obtain
\begin{align}
K= &\frac{n_e e^2}{mc}
\left[ 1 + \frac{1}{4}T\sum_{\omega_n} \int d\xi \textrm{Tr}
\left\{ \mathcal{G} \mathcal{G} + \mathcal{F}\underline{\mathcal{F}} \right\} 
\right],
\end{align}
at $m_1=m_2=m$, where $\xi=k^2/2m-\mu_F$, $n_e=4n_0$ with
$n_0$ being the electron density for each band per spin, and
$\mathrm{Tr}$ represents summation over spin and band degree of freedom.
We have used $N_0 v_F^2/d=n_0/m$ with 
$N_0$ being the density of states at the Fermi level in each band per spin.
Since the relation
\begin{align}
\frac{1}{4} \frac{1}{V_{\mathrm{vol}}} \sum_{\boldsymbol{k}}\left[  T\sum_{\omega_n} \frac{v_F^2}{d} \textrm{Tr}
\left\{ \mathcal{G}_N \mathcal{G}_N \right\} \right]= - \frac{n_0}{m},
\end{align}
holds for Green function in the normal state $\mathcal{G}_N$~\cite{agd}, the Meissner kernel becomes
\begin{align}
K= &\frac{n_e e^2}{mc} Q,\\
Q=&\frac{1}{4} T\sum_{\omega_n} \int d\xi \textrm{Tr}
\left\{ \mathcal{G} \mathcal{G} + \mathcal{F}\underline{\mathcal{F}} - \mathcal{G}_N \mathcal{G}_N\right\}, 
\end{align}
where $Q$ corresponds to the dimensionless pair density. 
In the simple model~(\ref{smodel}), we obtain $Q=1$ at the zero temperature.

\end{document}